\documentclass{PoS}

\usepackage{amsmath} 
\usepackage{graphicx}
\usepackage{amssymb}
\usepackage{axodraw}

\PoS{PoS(LAT2005)234}

\title{Dimensional regularization of Schr\"odinger Functional correlation
functions\footnote{FTUAM-05-14 and IFT-UAM/CSIC-05-38}}

\ShortTitle{Dimensional regularization of Schr\"odinger Functional
correlation functions}

\author{Eduardo Obeso\\
Departamento de F\'isica Te\'orica C-XI and Instituto de F\'isica Te\'orica C-XVI,\\
Universidad Aut\'onoma de Madrid, Cantoblanco, 28049 Madrid, Spain\\
        E-mail: \email{eduardo.obeso@uam.es}}

\abstract{The matching between Schr\"odinger Functional renormalization schemes and conventional perturbative schemes is usually done using an intermediate lattice scheme. We propose to do the matching directly. This requires the perturbative evaluation of Schr\"odinger Functional correlation functions in the continuum. We use dimensional regularization but due to the lack of translational invariance in the Euclidean time direction, we employ a general technique introduced by L\"uscher. In this talk I describe this technique and its application to the one-loop expansion of correlation functions used in the definition of the renormalized quark mass in the Schr\"odinger Functional scheme. The divergent parts are identified and the computation of finite parts is in progress.}

\FullConference{XXIIIrd International Symposium on Lattice Field
Theory\\
		 25-30 July 2005\\
		 Trinity College, Dublin, Ireland}

\begin{document}


\newcommand{\be}{\begin{equation}}
\newcommand{\ee}{\end{equation}}
\newcommand{\bm}{\begin{displaymath}}
\newcommand{\enm}{\end{displaymath}}

\newcommand{\slsh}{\slash \!\!\!}
 
\newcommand{\tderiv}{\partial_{0}}

\newcommand{\hx}{\hat x}
\newcommand{\hy}{\hat y}
\newcommand{\hmu}{\hat{\mu}}
\newcommand{\hnu}{\hat{\nu}}

\newcommand{\hbx}{\hat{\bf x}}
\newcommand{\hby}{\hat{\bf y}}
\newcommand{\hbz}{\hat{\bf z}}
\newcommand{\hp}{\hat{\bf p}}

\newcommand{\psibar}{\overline{\Psi}}


\newcommand{\ms}{\overline{\textrm{MS}}}

\newcommand{\momtheta}{\frac{\mathbf{\theta}}{L}}

\newcommand{\Tr}{{\rm Tr}}
\newcommand{\tr}{{\rm tr}}

\section{Introduction}
The QCD Schr\"odinger Functional (SF) \cite{SF,Stef1} can be used to define finite volume renormalization schemes which are suitable for non-perturbative renormalization. This means that the running of renormalized parameters and operators can be traced non-perturbatively from low to high energies through numerical simulations. Applications include the running coupling and quark mass, moments of structure functions, the static light axial current and four-quark operators (see \cite{4quarksop} and references therein for previous applications). The perturbative, high energy matching with conventional perturbative schemes such as $\ms$, is usually established by using an intermediate lattice renormalization scheme. It is possible, however, to do this matching directly through dimensional regularization of the SF. This is the aim of our work. Proceeding this way we expect advantages for the renormalization of four-quark operators and useful checks of previous results.

Conventional dimensional regularization techniques rely on the form of propagators in momentum space. Due to Dirichlet boundary conditions in the Euclidean time direction the standard techniques cannot be applied to the SF. Therefore, we will employ a  method due to L\"uscher \cite{Luscher}, using the heat kernels representation of propagators. As a first exercise we study the renormalization of the quark mass in the SF renormalized scheme \cite{Jansen}.

\section{SF renormalized quark mass}

We consider QCD with two flavors of mass degenerate quarks. The starting point for the definition of the SF renormalized mass is the PCAC relation:
\begin{equation}
\partial_{\mu}A_{\mu}^{a}(x)=2mP^{a}(x),
\end{equation}
where $A_{\mu}^{a}(x)$ and $P^{a}(x)$ are the isovector axial current and pseudoscalar density respectively. This relation holds true, up to contact terms, when inserted in any correlation function. The axial current is protected against renormalization by chiral symmetry so it stays finite. Thus if we define the renormalized mass $m_{R}=Z_{m}m$ through the PCAC relation, its renormalization will be given once an independent renormalization condition is imposed on the pseudoscalar density, i.e.~$Z_{m}=Z_{P}^{-1}$ given $P^{a}_{R}(x)=Z_{P}P^{a}(x)$. The renormalized pseudoscalar density is defined within the SF framework.

The SF geometry consists in a space-time manifold with Dirichlet boundary conditions in the Euclidean time direction and periodic boundary conditions (up to a phase for fermionic fields\footnote{This phase plays no r\^ole in the following discussion and will be omitted to ease the notation.}) in the spatial directions. To prepare the ground for dimensional regularization we consider a $D$-dimensional manifold with an arbitrary number $3+d$ of L-periodic spatial directions. Our notational conventions for indices and vectors are the same as in \cite{Stef2}. External momenta have only physical components and we take the boundary fields to be independent of the extra-dimensional coordinates. We impose vanishing gauge fields at the Euclidean time boundaries so that the induced  background field vanishes. 

To define $Z_{P}$ within the SF framework we consider bilinear source fields at the boundaries $x_0=0$ and $x_0=T$:
\be
\mathcal{O}^a\equiv\int d^{D-1}
\hat{\bf{y}}\;d^{D-1}\hat{\bf{z}}\;\overline\zeta(\hat{\bf{y}})\gamma_{5}
\frac{1}{2}\tau^{a}\zeta(\hat{\bf{z}}),\hspace{1cm}
\mathcal{O}'^a\equiv\int d^{D-1}
\hat{\bf{y}}\;d^{D-1}\hat{\bf{z}}\;\overline\zeta'(\hat{\bf{y}})\gamma_{5}
\frac{1}{2}\tau^{a}\zeta'(\hat{\bf{z}}),
\ee
where $\zeta$, $\overline\zeta$, $\zeta'$ and $\overline\zeta'$ are the usual boundary quark fields \cite{Stef22} and $\tau^{a}$ are Pauli matrices acting in flavor space. Boundary fields renormalize multiplicatively with a  common renormalization constant $Z_{\zeta}$ \cite{Stef2,Stef22}. We now introduce the following correlation functions:
\begin{equation}
f_{P}(x_0)=-\frac{1}{3}\langle P^{a}(x)\mathcal{O}^{a}\rangle,\hspace{1cm}f_{1}=-\frac{1}{3}L^{-2(D-1)}\langle \mathcal{O}'^{a}\mathcal{O}^{a}\rangle,
\end{equation}
where the powers of $L$ are such that the correlation functions are dimensionless. We write their perturbative expansion on the renormalized coupling $g_{R}$ as:
\be
f_{P}(x_0)=\sum_{n=0}^{\infty}f_{P}^{(n)}(x_0)g_{R}^{2n},\hspace{1cm}f_{1}=\sum_{n=0}^{\infty}f_{1}^{(n)}g_{R}^{2n}.
\ee 
 Now $Z_{P}$ is defined equating a ratio of renormalized correlation functions to its tree level value\cite{Jansen}: 
\be
\frac{f_{P}(x_0)_{R}}{\sqrt{(f_1)_{R}}}\equiv\frac{Z_{P}Z_{\zeta}^{2}f_{P}(x_0)}{\sqrt{Z_{\zeta}^{4}f_1}}\equiv Z_{P}\frac{f_{P}(x_0)}{\sqrt{f_1}}=\frac{f_{P}^{(0)}(x_0)}{\sqrt{f_1^{(0)}}},
\ee
where correlations are computed in the chiral limit and $x_0=T/2$. To establish the matching  between the SF and the $\ms$ renormalized mass we must consider the perturbative expansion:

\be
Z_{m}=\sum_{n=0}^{\infty}Z_{m}^{(n)}g_{R}^{2n}= 1+g_{R}^2\left(\frac{f_{P}^{(1)}(x_0)}{f_{P}^{(0)}(x_0)}-\frac{f_{1}^{(1)}}{2f_{1}^{(0)}}\right)+O(g_{R}^4).
\label{expansion}
\ee
Here we have used $Z_{m}=Z_{P}^{-1}$. In the following section we will see how to compute $f_{P}^{(1)}$ and $f_{1}^{(1)}$ using dimensional regularization.

\section{One-loop diagrams}

It is not difficult to obtain the perturbative expansion of our correlation functions. At order $g_{R}^2$ we find the following diagrams:

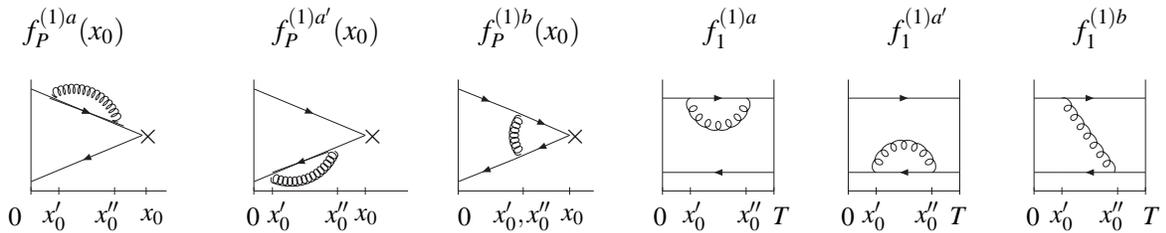
\begin{picture}(120,100)(0,0)    

\SetScale{0.70}

\Line(-20,85)(-20,80)
\Line(-20,25)(-20,30)
\SetColor{Black}          
\ArrowLine(-20,80)(40,55)
\SetColor{Black}
\ArrowLine(-10,75)(30,60)
\SetColor{Black}
\ArrowLine(40,55)(-20,30)
\Line(-20,80)(-20,30)
\SetColor{Black}
\GlueArc(3,55)(25,118,17){-2.5}{15}
\put(26,35.5){$\times$}
\SetColor{Black}
\Line(-20,25)(53,25)
\put(-20,75){ $f^{(1) a}_{P}(x_0)$}
\Text(-20,8)[]{\small 0}
\Text(-5,8)[]{\small$x'_0$}
\Text(15,8)[]{\small$x''_0$}
\Text(32,8)[]{\small$x_0$}
\Line(-20,27)(-20,23)
\Line(-5,27)(-5,23)
\Line(25,27)(25,23)
\Line(42,27)(42,23)

\SetColor{Black}
\Line(100,85)(100,80)
\Line(100,25)(100,30)
\SetColor{Black}
\ArrowLine(100,80)(160,55)
\Line(160,55)(100,30)
\SetColor{Black}
\ArrowLine(140,46)(110,35)
\SetColor{Black}
\Line(100,80)(100,30)
\SetColor{Black}
\GlueArc(120,55)(25,-115,-20){2.5}{15}
\SetColor{Black}
\Line(100,25)(183,25)
\put(111,35.5){$\times$}
\put(77,75){$f^{(1)a'}_{P}(x_0)$}
\Text(70,8)[]{\small 0}
\Text(80,8)[]{\small$x'_0$}
\Text(102,8)[]{\small$x''_0$}
\Text(113,8)[]{\small$x_0$}
\Line(100,27)(100,23)
\Line(110,27)(110,23)
\Line(145,27)(145,23)
\Line(160,27)(160,23)

\SetColor{Black}
\Line(210,85)(210,80)
\Line(210,25)(210,30)
\SetColor{Black}
\ArrowLine(210,80)(240,68)
\ArrowLine(240,42)(210,30)
\Line(210,80)(210,30)
\SetColor{Black}
\ArrowLine(240,68)(270,55)
\ArrowLine(270,55)(240,42)
\GlueArc(255,55)(15,138,222){2.5}{7}
\put(187.5,35.5){$\times$}
\SetColor{Black}
\Line(210,25)(283,25)
\put(155,75){${f^{(1)b}_{P}(x_0)}$}
\Text(150,8)[]{\small 0}
\Text(172,8)[]{\small$x'_0,x''_0$}
\Text(191,8)[]{\small$x_0$}
\Line(210,27)(210,23)
\Line(245,27)(245,23)
\Line(270,27)(270,23)

\SetColor{Black}
\Line(320,85)(320,75)
\Line(320,25)(320,35)
\Line(380,85)(380,75)
\Line(380,25)(380,35)
\SetColor{Black}
\Line(320,75)(320,35)
\Line(380,75)(380,35)
\Line(320,75)(335,75)
\Line(365,75)(380,75)
\ArrowLine(380,35)(320,35)
\SetColor{Black}
\ArrowLine(335,75)(365,75)
\GlueArc(350,75)(15,-180,0){2.5}{7}
\SetColor{Black}
\Line(320,25)(380,25)
\put(240,75){$f^{(1)a}_{1}$}
\Text(224,8)[]{\small 0}
\Text(237,8)[]{\small$x'_0$}
\Text(258,8)[]{\small$x''_0$}
\Text(270,8)[]{\small$T$}
\Line(320,27)(320,23)
\Line(335,27)(335,23)
\Line(365,27)(365,23)
\Line(380,27)(380,23)

\SetColor{Black}
\Line(420,85)(420,75)
\Line(420,25)(420,35)
\Line(480,85)(480,75)
\Line(480,25)(480,35)
\SetColor{Black}
\Line(420,75)(420,35)
\Line(480,75)(480,35)
\Line(420,35)(435,35)
\Line(465,35)(480,35)
\ArrowLine(420,75)(480,75)
\SetColor{Black}
\ArrowLine(465,35)(435,35)
\GlueArc(450,35)(15,0,180){2.5}{7}
\SetColor{Black}
\Line(420,25)(480,25)
\put(310,75){$f^{(1)a'}_{1}$}
\Text(295,8)[]{\small 0}
\Text(305,8)[]{\small$x'_0$}
\Text(325,8)[]{\small$x''_0$}
\Text(337,8)[]{\small$T$}
\Line(420,27)(420,23)
\Line(435,27)(435,23)
\Line(465,27)(465,23)
\Line(480,27)(480,23)

\SetColor{Black}
\Line(520,85)(520,75)
\Line(520,25)(520,35)
\Line(580,85)(580,75)
\Line(580,25)(580,35)
\SetColor{Black}
\Line(520,75)(520,35)
\Line(580,75)(580,35)
\Line(520,35)(535,35)
\Line(565,35)(580,35)
\ArrowLine(520,75)(580,75)
\SetColor{Black}
\ArrowLine(565,35)(535,35)
\Gluon(535,75)(563,35){2.5}{7}
\SetColor{Black}
\Line(520,25)(580,25)
\put(380,75){$f^{(1)b}_{1}$}
\Text(365,8)[]{\small 0}
\Text(376,8)[]{\small$x'_0$}
\Text(395,8)[]{\small$x''_0$}
\Text(410,8)[]{\small$T$}
\Line(520,27)(520,23)
\Line(535,27)(535,23)
\Line(565,27)(565,23)
\Line(580,27)(580,23)

\end{picture}

Arrows represent free quark propagators and curved lines represent free gluon propagators. The pseudoscalar density insertion is marked with a cross and vertical lines are Euclidean time boundaries. Self-energy and vertex diagrams are divergent. They can be compactly written like:
\begin{align}
\Sigma(f)&\equiv\int_{0}^{T}
dx'_0\,dx''_0\;\Tr\{f(x'_0,x''_0)\Sigma(x'_0,x''_0)\},\nonumber\\
V(f)&\equiv\int_0^{T}dx'_0\;dx''_0\;\Tr\{f(x'_0,x''_0)V(x_0,x'_0,x''_0)\}.
\label{vdiagram}
\end{align}
Here $\Sigma(x'_0,x''_0)$ and $V_P(x_0,x'_0,x''_0)$ are the amputated self-energy and vertex diagram respectively and they contain a sum over momenta. Different self-energy diagrams, $a$ and $a'$, are obtained changing the external function $f(x'_0,x''_0)$, which is a string of free propagators and gamma matrices.

\subsection{Dimensional regularization technique}

To regularize the one-loop diagrams we employ a general dimensional regularization technique as introduced by L\"uscher in \cite{Luscher}. We begin with a suitable representation for free quark and gluon propagators  similar to the one given in \cite{Stef2}, except that here we work in a time-momentum representation:
\begin{align}
\hat S^{(0)}(x_0,y_0;\hat{\bf{p}})&=[-\partial_0\gamma_0-ip_{\hat k}\gamma_{\hat{k}}][P_{+}\hat{G}^{ND}(x_0,y_0;\hat{\bf{p}})+P_{-}\hat{G}^{DN}(x_0,y_0;\hat{\bf{p}})]\nonumber\\
\hat D_{\hat\mu\hat\nu}^{(0)}(x_0,y_0;\hat{\bf{p}})&=\delta_{\hat{\mu}0}\delta_{\hat{\nu}0}\hat{G}^{NN}(x_0,y_0;\hat{\bf{p}})+\delta_{\hat\mu\hat{k}}\delta_{\hat{\nu}\hat{k}}\hat{G}^{DD}(x_0,y_0;\hat{\bf{p}}).
\end{align} 
Here $P_{\pm}=\frac{1}{2}(1\pm\gamma_0)$ are Dirac space projectors  and $G^{B_{0}B_{T}}(x_0,y_0;\hat{\bf{p}})$ are Green functions of the extended Laplacian operator, $\triangle=-\partial_{0}^{2}+\hat{\bf{p}}^2$, with Dirichlet (D) or Neumann (N) conditions at $x_0=0$ and $x_0=T$. The advantage of this representation is that now we can express the Green functions in terms of heat kernels:
\be
\hat{G}^{B_{0}B_{T}}(x_0,y_0;\hat{\bf{p}})=\int_0^{\infty}dt\;\hat{K}^{B_{0}B_{T}}_t(x_0,y_0;\hat{\bf{p}}),
\ee
where $\hat{K}_{t}^{B_{0}B_{T}}(x_0,y_0;\hat{\bf{p}})$ satisfy the heat equation:
\be
(\partial_{t}+\triangle)\hat{K}^{B_0B_T}_{t}(x_0,y_0;\hat{\bf{p}})=0.
\ee
Contributions from different dimensions factorize, in particular we can take out a factor which depends on extra, non-physical, components of the momentum:
\be
\hat{K}_{t}^{B_0B_T}(x_0,y_0;\hat{\bf{p}})=K_{t}^{B_0B_T}(x_0,y_0;{\bf{p}})\prod_{\mu=4}^{D-1} e^{-tp_{\mu}^{2}}.
\ee
Here $K_{t}^{B_0B_T}(x_0,y_0,\hat{\bf{p}})$ is the heat kernel of the Laplacian in 4 dimensions with the given b.c.

For illustration we will consider the vertex diagram. Factorizing the contributions from extra components of the momentum it can be written in the form:
\be
\begin{split}
&V(f)=\int_{0}^{\infty}dt_1\;dt_2\;dt_3\;[r(t_1+t_2+t_3)]^{d}\;I(f;t_1,t_2,t_3),
\\&\textrm{with:}\hspace{1cm}r(t)\equiv L^{-1}\sum_{n=-\infty}^{\infty}e^{-tp^2(n)},\hspace{1cm}p(n)\equiv\frac{2\pi n}{L}.\end{split}
\label{vdpropertimes}
\ee
It is at this point that we can take the number of extra dimensions $d$ to be a complex number $d=-2\epsilon$, and the diagram becomes a meromorphic function of $\epsilon$. At $\epsilon=0$ the integration diverges due to singularities in the integrand $I(f;t_1,t_2,t_3)$ when all proper times go to zero simultaneously. Since the function $r(t)$ goes as $(4\pi t)^{-1/2}$ for small t, the prefactor $[r(t_1+t_2+t_3)]^{-2\epsilon}$ acts as a regulator and cancels the divergences of the integrand when $Re(\epsilon)$ is large enough. The diagram at $\epsilon=0$ must be defined by analytical continuation as we will see in the following subsection.
\subsection{Analytical continuation to $\epsilon=0$}

We now look in some detail at the integrand: 
\begin{align}
I(f;t_1,t_2,t_3)&=-C_{F}L^{-3}\int_{0}^{T}dx'_{0}dx''_{0}\sum_{\bf{p}}e^{-(t_1+t_2+t_3)\bf{p}^2}\Tr\left\{\gamma_{\hat{k}}\left[\Gamma^{A}\partial_{0}K_{t_1}^{ND}(x'_0,x_0)\partial_{0}K_{t_2}^{DN}(x_0,x''_0)-\right.\right.\nonumber\\&\left.\left.-{\bf{p}^2}\Gamma^{B}K_{t_1}^{ND}(x'_0,x_0)K_{t_2}^{ND}(x_0,x''_0)\right]K_{t_3}^{DD}(x'_0,x''_0)\gamma_{\hat{k}}f(x'_0,x''_0)\right\}+\cdots
\end{align}
 Here $C_{F}=(N^2-1)/2N$  with $N$ the number of colors. Omitted terms have the same structure. We have factorized the kernels for the 3 spatial directions and the one-dimensional kernels  for the Euclidean time direction: $K^{B_0B_T}_{t}(x_0,y_0)$. An explicit expression for them can be found in \cite{Stef2}. $\Gamma^{A}$ and $\Gamma^{B}$ are some Dirac structures. To analytically continue to $\epsilon=0$ we first must identify at which values of proper times this expression is not integrable in the absence of regulator. Singularities arise at small values of proper times. We study the asymptotic behavior of the integrand when one, two, or all proper times go to zero and find that, as anticipated in the previous section, only in the latter case we obtain a non integrable expression. It is convenient then to do the following change of variables: $t_1=ts_1$, $t_2=ts_2$, $t_3=ts_3$, with $s_1+s_2+s_3=1$ and the vertex diagram can be written:
\be
V(f)=\int_{0}^{\infty}dt\;t^2[r(t)]^{-2\epsilon}\int_{0}^{1}ds_1\;ds_2\;ds_3\;\delta(s_1+s_2+s_3-1)I(f,ts_1,ts_2,ts_3).
\label{diag3}
\ee
To analytical continue to $\epsilon=0$ we need to know the asymptotic expansion of $I(f,ts_1,ts_2,ts_3)$ including terms of order $t^{-3}$. Since exponentially suppressed terms do not enter in the asymptotic expansion, we can consider reduced heat kernels (see \cite{StefTh}). Furthermore,  for the vertex diagram it is found that only the free space  contribution, i.e.~ the one  we would obtain if the boundaries were not present, contributes to the divergence. We can then, from the very beginning of the calculation, split propagators and the corresponding kernels in its free space part (f) and  a surface part (s) \cite{Symanzik}:
\be
S=S^{f}+S^{s},\hspace{1.2cm}D_{\hmu\hnu}=D_{\hmu\hnu}^{f}+D_{\hmu\hnu}^{s}.
\ee
Contributions to the diagram with any surface propagator are finite and can be computed without introducing the heat kernel representation. For the contribution coming from considering the free part for all propagators we use heat kernels and find:
\be
I^{f}(f;ts_1,ts_2,ts_3)\stackrel{t\rightarrow0}{\sim}A(f;s_1,s_2,s_3)t^{-3}.
\ee
$A(f;s_1,s_2,s_3)$ is a functional of the external function $f$, well behaved in the domain of integration. The superscript $f$ reminds us that we are working only with free kernels. Since the leading term goes like $t^{-3}$, it is not necessary to go further in the asymptotic expansion. Inserting this expression in Eq.~(\ref{diag3}) and using the asymptotic behavior of the regulator, we perform the integration over proper times at the lower end of the domain of integration  and obtain the divergence as a pole in $\epsilon$. To make the analytical continuation to $\epsilon=0$ we add and subtract this divergence and obtain the following representation of the diagram:
\begin{align}
V^{f}(f)&\stackrel{\epsilon\rightarrow 0}{=}\int_{0}^{1}ds_1\;ds_2\;ds_3\;\delta(1-s_1-s_2-s_3) \left\{A(f;s_1,s_2,s_3)\left[\frac{1}{\epsilon}+\ln(4\pi)\right]+\right.\nonumber\\
&\left.+\int_{0}^{\infty}dt\;t^2\left[I^{f}(f,ts_1,ts_2,ts_3)-\theta(1-t)A(f;s_1,s_2,s_3)t^{-3}\right]\right\}+O(\epsilon).
\label{analyticc}
\end{align}
The first term is the divergence and the remaining terms are finite contributions. Proceeding analogously with the self-energy diagram we obtain the following divergences:
\begin{align}
f_{P}^{(1)}(x_0)|_{\textrm{pole}}&=\frac{-C_{F}}{4\pi^2\epsilon}f_{P}^{(0)}(x_0)+\frac{C_{F}}{4\pi^{2}\epsilon}f_{P}^{(0)}(x_0)=0,\\
f_{1}^{(1)}|_{\textrm{pole}}&=\frac{-3C_{F}}{8\pi^2\epsilon}f_{1}^{(0)}.
\end{align}
The first term in the sum comes from the self-energy and the second from the vertex diagram. Introducing these results in the expansion of Eq.~(\ref{expansion}) we find the correct divergence for the quark mass:
\be
Z_{m}^{(1)}|_{\textrm{pole}}=\frac{3C_{F}}{16\pi^2\epsilon}. 
\ee

\section{Conclusions and final remarks}

We have seen how to apply dimensional regularization to SF correlation functions. In particular we have identified the divergences of the one-loop expansion of the correlation functions used to define the SF renormalized quark mass. However it is the finite part what is required to establish the matching with other perturbative schemes.

The computation of the finite part can be divided in different pieces. First of all, the free gluon propagator has a particular expression for zero momentum. We must compute the zero momentum contribution to our diagrams separately then. This has been obtained analytically using MAPLE. On the other hand, for the computation of the divergence only reduced kernels were required, in particular for the vertex diagram only the free part contributes. We can compute the remaining contribution working with $S^{s}$ and $D_{\hmu\hnu}^{s}$, which have a simpler analytical expression. Here only a sum over momenta and two integrations are required. Finally we have the finite contributions coming from Eq.~(\ref{analyticc}). Obviously the difficult part here comes from the subtraction of the divergence from the integrand. There are five integrations and the sum over momenta to be done. All these computations are work in progress.

 {\bf{Acknowledgements:}} I would like to thank S. Sint for numerous helpful discussions and comments on the manuscript. This work is supported by the Ministerio de Ciencia y Tecnolog\'ia (Spain) through a FPI grant.

\end{document}